# Multimodal Surrogates for Video Browsing

Wei Ding, Gary Marchionini, and Dagobert Soergel
University of Maryland at College Park & University of North Carolina at Chapel Hill
{weid@glue, ds52@umail}.umd.edu, march@ils.unc.edu

**Abstract**
Three types of video surrogates--visual (keyframes), verbal (keywords/phrases), and combination of the two--were designed and studied in a qualitative investigation of user cognitive processes. The results favor the combined surrogates in which verbal information and images reinforce each other, lead to better comprehension, and may actually require less processing time. The results also highlight image features users found most helpful. These findings will inform the interface design and video representation for video retrieval and browsing.

**Introduction**

Use of digital videos on the Web is becoming more common for education and academic communication as well as for entertainment. As video collections grow, users must be provided with effective searching and browsing tools to facilitate quick and easy access. This requires appropriate surrogates to represent original video documents.

Surrogates (e.g., bibliographic citations, abstracts, and table of contents) are used in most information retrieval systems. Browsing surrogates allows users to make quick decisions about whether to examine information objects in greater detail--result examination) and supports incidental learning (users can capture the most interesting information without reading/viewing everything in full size)--information extraction. Surrogates also support faster lookups and network transfers.

Creation and evaluation of surrogates are long-standing research issues (Borko & Bernier, 1975). The nature of video data and the active user needs for video bring new challenging issues related to the design and display of video surrogates. Because pictures, motion, speech, and other audio all communicate important information in video, purely linguistic approaches to indexing and representing videos are generally agreed to be insufficient. Image-based approaches to surrogate creation are also required. For the image sequences in video, surrogates have been suggested such as keyframes (e.g., O'Connor, 1985; Zhang, et al., 1995) and salient stills (e.g., Teodosio & Bender, 1993). These approaches use visual or audio signal processing (e.g., color, luminosity, optical flow, texture) to detect features that might be used to cue human recognition and recall. Such techniques have attracted many researchers because of the feasibility of automatic processing and the self-evidence of pictures (impart information that cannot be expressed via words). Nevertheless, it is not yet evident which is the best approach to representing videos without examining the underlying cognitive judgments that users bring to the representations.

Studies into video and still images suggest that words could be as powerful as pictures for communication. Cawkell (1995) predicts that pictures will not always outperform words because they do not always replace the descriptive power of words that may be better for some abstract concepts. It is a general requirement to supply textual descriptors in order to identify specifics in all commercial still and moving image documents (Enser, 1995). When one video clip is represented by multiple still images, processing of image sequences by users will involve some additional issues because the meaning of words as signs are generally agreed on and made more specific by syntax, whereas pictures are specific and made general by their context (Pryluck, 1976). Therefore, interpretation of visual surrogates could be more ambiguous than verbal ones. O'Connor (1991), the first researcher proposing using key frames as video surrogates, also noted that still images represent only one small fragment of the time continuum represented by the moving image document. Some images together with some words may well be adequate to guide a user among many documents on similar topics. Turner (1994) posits that for access to video information, text and images are complementary and interdependent. A series of studies conducted by the authors and colleagues (summarized in Tse et al, 1998) with video surrogates composed of keyframes show that keyframes only (as surrogates) could well support rapid identification of visual objects, but the comprehension of the video story was rather limited.

According to the research on information objects involving more than one modality (e.g., full motion videos), reinforcing rather than interfering effects can be expected once the information is well integrated. Redundancy theory suggests that the redundant information from different modalities provide cross-references to the target to be understood (Pryluck, 1976), overcome the limitations of individual information channels (Morgan & Welton, 1992), and increase possibilities for comprehension (Stone & Gluck, 1980). Based on brain lateralization and parallel processing theories, Kantowitz (1985) hypothesized that redundant information simultaneously perceived through two channels (images and words) actually speeds up processing time. However, little empirical study has been done to test whether these findings on primary information objects are applicable to surrogates.

This study is a follow-up of a quantitative experiment (Ding, in progress), which investigated the information representation power of different modalities in video data based on user performance (accuracy and response time) and user preference. Both studies are built upon the Baltimore Learning Community project (BLC)(Marchionini et al, 1997), which allows public school teachers to access a web-based multimedia database and link relevant instructional resources to their lesson plans and class representations. Included in this digital library are still images, websites, and texts, and segmented educational documentary videos from the Discovery Channel, Maryland Public Television, and the National Archives. The interface provides both visual (keyframes) and verbal (bibliographic information and abstracts) surrogates as video previews.

The goals of this study were to better understand the roles of visual and verbal information in representing multimedia documents, to reveal the cognitive processes involved in video surrogate examination more comprehensively, and to identify various decision patterns and impact factors behind the user behavior.

**Research Questions**

Focusing on people's cognitive processing and reasoning mechanisms involved in video surrogate examination, this study aims to answer a series of research questions organized around three aspects: the usefulness of the multimodal (combined)surrogates, the decision-making processes in surrogate examination, and the mechanisms in video comprehension.

- Usefulness of the combined surrogates

What video surrogates did participants prefer? Why? Why did participants like combined surrogates with keyframes and keywords/phrase? What did verbal information and visual information each contribute to the surrogate? What time-accuracy tradeoffs were involved?

- Decision making processes

What decision making processes were involved in the visual gisting and verbal comprehension tasks? How might understanding these processes inform the design of video representations?

- Video comprehension

How did people describe/make sense of videos? Did they treat all the available information equally or not? What kind of information was more useful and attention capturing? How did participants take advantages of those kinds of information?

**Related Work**

Compared to the effort involved in the content-based video indexing and retrieval technology, the community has witnessed a dearth of evaluation studies with users' participation. How to leverage the available video visualization techniques and accommodate user needs into effective interfaces for video browsing and access continues to be an important issue.

Despite the strong research interest in factors of users and information for relevance judgments (e.g., Wang & Soergel, 1998), studies particularly focusing on video documents are rather sparse. Goodrum (1997) compared four different video surrogates (representations) (key frames, salient still, title and key words) for 12 10-second video clips (without sound) from the CNN environmental unit, and found image-based representations support higher congruence in similarity judgments than do text-based (title and key words) representations. For specific queries, text-based representations force higher congruence in utility judgments than do the image-based ones. The hypothesis that image-based representations have advantages over text-based ones for generic queries was not supported. Christel et al. (1997) compared different video result presentations (poster frame menu vs. text title menu), and found that poster frames, when chosen based on the query, lead to significantly faster location of the relevant video (facts-finding) by a user over the presentation of only a text title menu.

An earlier study investigated the pattern of information uses for video selection in manual video libraries (Cohen, 1987). The results highlight the importance of subject (topicality) information for video selection.

**Methodology**

This study took an exploratory approach as a follow-up to the quantitative study that showed that users preferred combined surrogates while performing similarly with image-only, text-only, and combined surrogates (Ding, in progress). Data collection was mainly through observations and user's talking aloud while performing recognition and comprehension tasks.

Experiment material

Fourteen 2-3 minute video clips were selected from a collection of 24 one-hour Discovery documentaries in the BLC database. For each video clip, three types of video surrogates were created.
- Verbal surrogate-- Six keywords/phrases were manually picked from the audio channel, or assigned based on the overall meaning of the video.
- Visual surrogate-- Organized as a storyboard, the twelve best keyframes were selected from the keyframes automatically extracted by a scene-change based segmentation program MERIT (Kobla et al., 1997).
- Combined surrogate-- Six keywords/phrases were listed at the top of the storyboard of 12 keyframes.

User tasks

We developed two user tasks--verbal comprehension and visual gisting, which are needed in authentic information seeking activities for quick video browsing and result examination in a real video database. These tasks and techniques to measure them have been used in several previous studies and refined here (summarized in Tse et al, 1998).

*Verbal Comprehension*

Verbal comprehension is the extent to which the user can get the main idea of the video clip from the surrogates. Better comprehension would allow more accurate relevance judgments and video selections. There were two sub-tasks: free-form writing, and selection of phrases/sentences. For the writing task, participants were asked to write 2-3 sentences summarizing what the video clip was about. For the true/false judgment task, there were 6 summary statements /phrases for each clip, some correct and some not (in random proportion and sequence). The statements were shown on the screen one at a time, with participants being asked to make judgments by clicking on the corresponding button. The testing statements were based on the transcripts, answers from the pilot participants, and the abstracts available in the Baltimore Learning Community (BLC) digital library.

*Visual gisting*

The purpose of the visual gisting task was to test the surrogate's information representation power non-verbally. It investigated to what extent users could perform "visual closure" by watching surrogates: A surrogate carries essential information of the video, and leaves blanks and uncertainties as well. When viewing the surrogate, users need to fill in the blanks and imagine what this video would look like. Participants were shown 10 test images, some from the video and some not (distractors) (in random proportion and sequence), one at a time, and asked whether the image belongs to the video. The test images did include images from the surrogate. Distractors were selected from other videos by the researcher and another BLC staff member and approved by a panel. This task allow s users to demonstrate comprehension through images rather than through linguistic devices, but participants in the textual surrogate group have to think across media.

Sampling of participants

This experiment took a "purposeful sampling" approach (Patton, 1990, p. 169), Participants were selected deliberately in order to provide important information that can't be gotten as well from other choices (Maxwell, 1996). Twelve volunteers were recruited: 4 schoolteachers, 4 graduate students in education (teachers to be), and 4 other graduate students with different majors (computer science, law, audiology, and biology).

Data Collection Methods

Think aloud (Ericsson and Simon, 1993) was adopted as the major method for data collection supplemented by observation and post hoc interviews. Different methods could supplement each other and overcome some deficiencies. For example, through observation and participant's think aloud, the researcher was able to compare what was said and what was done by the participant so that the subjectivity of verbalization could be minimized, information that was implicit or not recordable could be captured. Through the post hoc interview, misunderstandings or confusion can be clarified or dismissed. All the performance data were tracked by computer, and all sessions for thinking aloud and interviews were audio-tape recorded.

Experimental procedure

The study was conducted at the University of Maryland from July to October, 1998. Participants came to the researcher's office using a same computer (PC, Pentium 200, 17-inch monitor with resolution of 1024x768). Participants were randomly assigned to either the visual gisting task or the verbal comprehension task. After a practice session, the participant would be exposed to every surrogate treatment (keywords, keyframes, and the combination) in a random sequence, each with two sets of video surrogates. For each set of surrogates, the participant talked aloud what s/he saw in the surrogate, what the video was about, then proceeded to the user task. S/he would click on yes/no button to indicate whether the test picture belonged to the video (visual gisting) or the verbal statement correctly reflected the meaning of the video while talking aloud the decision steps. Participants were instructed to speak out "everything that runs across your mind"

Data analysis

Transcribed audio tapes and field notes were jointly coded and analyzed using NUD.IST (a program for analysis of non-structured data). Results and Discussion This paper focuses on the first research question, the usefulness of the combined surrogates and the value visual and verbal modalities bring to the surrogates. For details about other questions, please refer to the dissertation (Ding, in progress).

Usefulness of combined surrogates

All participants except one found the combined surrogates more useful than the single-modal ones, and most of them said "It depends" when asked about their preference between the other two types of surrogates. Words and images each provided unique information and served different purposes that might not be provided otherwise; words and images reinforced each other, facilitated information integration and video sense making. In addition, some users are more visually oriented while others are more verbally oriented; combined surrogates can meet the needs of both.

- Words tell the "aboutness" or the meaning of the image sequence (keyframe storyboard) and supplement the images.

Although participants could get some basic idea of what the video is about from the keyframes, which might be so broad as to fit into any subject matter, keywords often explicitly demonstrated the subject matter itself. For example, several participants mentioned in examining a combined surrogate (for a particular video), "Without the words I would have just seen a bunch of birds. Keywords tell a little bit about the story." Further, when there was a conflict between the images and words/phrases, participants tended to depend more on the words.

Uncertainty about visual details may hinder users from interpreting the video in its own way. Participants tended to ignore the information that is fuzzy, uncertain or confusing to them, and draw conclusions only based on information they were sure about. The video "Dry Season Animal Survival Strategy" describes how birds and monkeys share food during the dry season and their roles in the food chain in the rainforest. One participant, based on the keyframe surrogate, could not recognize the animal as a monkey, or realize it was a pollinator for the flowering trees. So he ignored the part of the monkey in the video, and thus missed the theme about "food chain" and "share". His conclusion was drawn based only on the activities of birds. In contrast, participants with the combined surrogate captured more themes of the video, and figured out more details from the keyframes. A more accurate and sound inference was made when the participant saw the keywords "food chain", "sharing", "flowering trees", and "monkey".

It is likely that the verbal information led to the construction of a story scheme which participants used to make sense of the video. The visual surrogates provided specific details for the story scheme. Also, the verbal information appeared to serve as a label for images or a container of the visual contents, which shaped the conceptual theme or outline, the main idea of the video.

- The uniqueness of keyframes and keyframe sequences

Although participants frequently mentioned the importance of verbal information in making sense of the videos, often the usefulness of the verbal information was built upon the key frames. The words/phrases facilitated the understanding of the keyframes and the keyframe sequence. In addition, participants commented that the uniqueness of images, such as concretion, vividness, impressiveness, and realism, cannot be easily or effectively be substituted by words. Images illustrated abstract concepts by providing "what it looks like"; They gave detailed information including settings, emotion and background as well as the main focus; Images were more interesting and fun to watch, and welcomed exploration and association. They could also be repurposed for multiple uses and interpretations. A teacher participant who strongly preferred keyframe surrogates to the others mentioned:

> Sometimes when you have the images, you don't have to focus on what the actual topic is. You can put it in another setting or create another setting with those images yourself, so that would give you a sense. Oh, I could use them this way, and I would have never thought about using it this way if I had heard the whole clip or the sound.

- Words/phrases and key frames reinforce each other.

The semantics of verbal information and the uniqueness of images reinforce each other. Participants explained why the combined surrogate was superior to the single modal surrogates from different perspectives: In terms of information sufficiency, text only and images only both constrained in fully understanding the video content. Putting them together could optimize the structure of the representation in that words and images would provide different kinds of information and support each other. From a user's perspective, the combined surrogate could better accommodate different user needs or different types of users. From a cost-effectiveness perspective, well-integrated surrogates allow effective examination per time investment.

- Text in images drew special attention.

Text in images (e.g., images with captions or graphics) was not studied as a specific surrogate type in this research. However, the key frames containing drew participants' attention due to the fact that they conveyed more information than regular keyframes. Two kinds of text appeared in the keyframes in this study: subtitles and graphics. Subtitles are captions in a different language from what is spoken in the video, and graphics refers to the consecutive frames in a video, in which text and pictures are both integrative elements. Participants used phrases like conveying "a lot of information", "indicative", "crucial" and "attention capturing". Text in images may be so useful because it is well integrated and matched with the corresponding images, which might even go beyond the role of keywords/phrases in the combined surrogates. Participants did not have to match the text and images themselves, which sometimes may cause ambiguity or confusion otherwise.

- Preference of video surrogates

Participants liked the idea of quickly browsing video surrogates to get the gist before spending time and effort watching the full motion video for the purpose of screening and selection. All participants except one claimed that they preferred the combined surrogates because of the information abundance, accommodation of different user needs, and user characteristics. This further confirmed the results from the quantitative study. Furthermore, participants seemed to agree that processing the combined surrogates did not take any longer; on the contrary, sometimes it could speed up the sense making process.

Decision making process in video surrogate examination

An understanding of the information processing mechanisms involved in combined surrogate examination may explain why participants preferred the combined surrogates from a different point of view; also it would shed light on techniques for presenting video surrogates in real video browsing settings.

Participants employed different strategies to making sense of a video based on the combined surrogates. Some participants claimed that they just followed the sequence in which the information was organized in the surrogate. Some first looked at image sequence as a whole briefly, and then went to the words, and finally went back to examine the images individually and carefully. The others first read the words, and then went to the images. Behind the different surrogate processing strategies, several impact factors were involved.

First, a processing strategy was related to a participant's preference of modality. For example, Participant 7 contended that once the keywords were carefully and properly chosen, images may not be necessary as part of the video surrogate. Accordingly, he always went to the words first. Participant 11, on the other hand, highly preferred images to words. She said that her eyes were first drawn to the images when she processed the first combined surrogate, and she thought that was automatic.

Second, the processing strategy was also dependent on the viewer's familiarity with the information in the surrogate. When the topic was not familiar, participants tended to first resort to the verbal information. Although Participant 11 laid her eyes on the images first, she could not get much information from there, and had to switch to the keywords. So, in second trial she went to the keywords first. But she commented that for that one (the Cherokees), the words did not help much. In processing the second combined surrogate, Participant 2 also ended up going to the words first.

Third, another factor that could direct the processing is the visual appeal/attractiveness of images. It is possible that attractiveness resulted from familiarity, but novelty (e.g., special attention paid to unusual scenes), emotion or something else were involved too. (to be discussed in the next section).

The surrogate information processing strategy depends on factors such as user's modality orientation, the user's knowledge about the surrogate content, and the visual appeal in the surrogate. These factors were intertwined from situation to situation, and caused participants to take different steps to reach their goal contextually.

Regardless of whether they started with words or images, participants adopted two main generic processing strategies: sequential vs. selective. With the first strategy, participants basically followed the sequence as the information was pre-structured.

The second strategy was more dynamic and proactive. Participants first built up a quick scheme of a story, and then tested whether the scheme still held with further details or specificity fitted in. It seems that the verbal information was more likely to be better suited to scheme building, and images are better for confirmation and illustration.

Participants addressed different uses of the sequence of the images. When there was no additional information available except for the images, they had to make full use of the sequence. When there was verbal information besides the images, they tried to absorb the visual information based on the visual cues.

There seems to be a lot of verbal/visual integration taking place. Time was not recorded in this experiment, but from the first experiment, there was no significant difference in processing time between the keyframe surrogate and the combined surrogate. It again suggests that the information integration is faster and easier

Participants liked the combined surrogates because they could cross-examine the surrogates, and build up and further refine a scheme to make sense of the video. Whether going to the images or the words first was mainly dependent on the personal orientation to the modality. Most participants seemed to first quickly scan the words or image sequence as a whole, and then switched to the other modality. No matter whether the participant processed the information more sequentially or selectively, they tended to rely more on the words than the images to set up the baseline of the story in the video. The intertwining of the multiple impact factors could cause the participant to adapt strategies and tactics to each specific situation.

Sense making of video surrogates

This section answers two main questions: what information most captured participants' attention, and based on what information in the surrogates participants tried to comprehend the video.

*Attention-getting information*

Participants did not attend to all the stimuli equally, instead, something usually first captured his/her attention, especially when keyframes were available in the surrogate. With the verbal surrogate, participants in the verbal comprehension task tended to make up a sentence by including all the keywords provided or by paraphrasing them. With surrogates that included visual elements, participants often paid more attention to particular keyframes if they noted the keyframe contained text (described in the previous section), interactive information, symbolic scenes, and unusual scenes. In addition, human beings or other living objects captured more attention than still objects.

Participants showed special interest in keyframes with one of the following features:
- Text in pictures -- Captions or graphics gave "voice" to images;
- Interaction information -- Action scenes attracted more interest than static scenes;
- Symbols -- Icons and stereotypes attract attention and cued visual gisting;
- Novelty-- Visual images gained attention that might lead to inferences;
- Emotion-- Scenes that evoked strong emotional response attracted attention.

*Sense-making strategies*

Participants tended to use the most attention-capturing cues to build up the theme about the video, and then used the other information to reinforce, confirm, or adjust the story. From the way participants described/imagined what the video was (should be) about it can be seen that the comprehension was centered to human beings and their activities. Also they a tried to describe the video with as many specific terms as possible --they tried to be specific with the name, location, and means. This was consistent with results from the first experiment that iconographical concepts were more frequently used than pre-iconographical concepts in summarizing the full motion video.

Participants showed a strong people-orientation (consistent with Valva's findings, cited in Massey and Bender, 1996). They tended to make up a story of a video by putting particular person(s) seen in the video at the core although in many cases that might not be true. Maybe it was easier to come up with a story involving people. For example, one video actually shows how the tribe lives in harmony with nature, and how they keep animals as pets, but participants paid more attention to the people. "So my story about this video is how the boy spent his leisure time with animals in the village." In the video Early Trains and Railroads, there was only one picture of a man demonstrating how the first telegram worked. The story to a participant was "I guess that guy is a railroad timekeeper so that they could use the telegram to see whether or not the train is on time and to ensure there won't be accidents …".

Participants would target other contents if people were not available. "We seem to have lots of birds and tropical habitats. It doesn't seem to be about people." "This video describes the hidden city showing details of the exterior and interior. Other than that, there are no people here anywhere. Very deserted."

Specificity orientation requires that the viewer have sufficient prior knowledge. Providing proper verbal information (e.g., keywords/phrases) might compensate for the lack of knowledge. People orientation seems to be a plausible strategy to make sense of the video, especially when available information is limited (e.g., when participants dealt with video surrogates). Future keyframe extraction techniques may need to take these factors into consideration.

## Conclusion

Our study shows that users strongly prefer video surrogates that combine verbal information (text in a generic sense) and images. Each modality makes a unique contribution to the comprehension of a video, and in combination they reinforce each other. Verbal information helps users get the overall meaning of the video and specify or clarify the thematic information described in the visual surrogates, such as who, where, when and how. Put differently, verbal information conveys the iconographical meaning, and supports users understanding of the contents of images and the meaning they refer to. On the other hand, visual information is concrete, vivid, detailed, and more real; it is more apt to convey affect, emotion, and excitement and to draw attention. Often verbal information helps the user to extract more meaning from images. The combined surrogates integrated verbal and visual information and facilitate information processing so that it actually may take less time to process the larger amount of information present in combined surrogates as compared to purely verbal or purely image surrogates. Further studies could investigate whether short abstracts instead of keywords and/or the audio presentation of the verbal part would bring still greater benefits.

While image-based surrogates, particularly those that can be prepared automatically, have received much attention. the message to designers is clear: Provide video surrogates that integrate images and text. Text may be descriptive of the video as a whole or label a specific image; both kinds of text are helpful. In selecting images for surrogates, such as in key frame extraction, consider images that depict interaction, especially of people, images that evoke emotions, images that contain text or symbols, and images that are novel and attractive — these kinds of images seem to help people most in making sense of a video.

**Acknowledgement**

This research is supported by the US Department of Education Technology Challenge Grant (#R303A50051) to the Baltimore Learning Community Project. The authors thank the Discovery Communications for the use of the educational videos. Thanks are also due to the participants in this study.